\documentclass[a4paper]{article}
\usepackage{graphicx}% Include figure files
\usepackage{bm}% bold math

\newcommand{\sech}{\mathrm{sech}}
\newcommand{\eq}[1]{Eq.~(\ref{#1})}
\newcommand{\fig}[1]{Fig.~\ref{#1}}         
\newcommand{\iotabar}{{\mbox{$\iota\!\!$-}}}
\newcommand{\iotadot}{\dot{\iotabar}}
\newcommand{\iotaddot}{\ddot{\iotabar}}
\newcommand{\Bvec}{{\mathbf{B}}}

\newcommand{\kvec}{{\mathbf{k}}}

\newcommand{\xivec}{{\bm{\xi}}}
\newcommand{\mmax}{m_{\mathrm{max}}}
\newcommand{\mumax}{\mu_{\mathrm{max}}}
\newcommand{\mumin}{\mu_{\mathrm{min}}}
\newcommand{\etamax}{\eta_{\mathrm{max}}}
\newcommand{\gammamax}{\gamma_{\mathrm{max}}}
\newcommand{\kappamax}{\kappa_{\mathrm{max}}}

\newcommand{\rmu}{r_{\mu}}

\newcommand{\half}{{\textstyle{\mathrm{\frac{1}{2}}}}}
\newcommand{\quarter}{{\textstyle{\mathrm{\frac{1}{4}}}}}

\begin{document}
\setlength{\baselineskip}{4.30mm}

\normalsize
%\small

\twocolumn[
%-----------------------------------------------------------

\setlength{\baselineskip}{6.0mm}

%%%%%%%%    Title of paper is written here.                          
%%%%%%%%

\vspace*{0.60in}

\begin{center}
{\large\bf 
   Quantum chaos theory and the spectrum \\
   of ideal-MHD instabilities in toroidal plasmas
}

\vspace*{0.30in}

%%%%%%%%    Author(s) of paper is written here.                      
%%%%%%%%
%%%%%%%%    Presenting auther should be written first.       
%%%%%%%%

   DEWAR Robert L, N\"UHRENBERG Carolin$^{ 1)}$ 
   and TATSUNO Tomoya$^{ 2)}$

%%%%%%%%    Institution of author(s) is written here.               
%%%%%%%%

{\it 
   Research School of Physical Sciences and Engineering, The 
Australian National University, Canberra ACT 0200, Australia  \\
   1) Max-Planck-Institut f\"ur Plasmaphysik, Teilinstitut Greifswald 
IPP-EURATOM Association, D-17489, Greifswald, Germany \\
   2) Institute for Research in Electronics and Applied Physics 
University of Maryland, College Park, MD 20742-3511, USA
}

%%%%%%%%    E-mail address of the presenting author.                 
%%%%%%%%

{\it 
   e-mail: robert.dewar@anu.edu.au
}

\end{center}

\vspace*{0.00in}

%%%%%%%%    Abstract                                                 
%%%%%%%%

\noindent
{\bf Abstract}
\setlength{\baselineskip}{4.30mm}

\hspace*{12pt} \small{ In a fully 3-D system such as a stellarator,
the toroidal mode number $n$ ceases to be a good quantum number---all
$n$s within a given mode family being coupled.  It is found that the
discrete spectrum of unstable ideal MHD (magnetohydrodynamic)
instabilities ceases to exist unless MHD is modified (regularized) by
introducing a short-perpendicular-wavelength cutoff.  Attempts to use
ray tracing to estimate the regularized MHD spectrum fail due to the
occurrence of chaotic ray trajectories.  In quantum chaos theory,
strong chaos in the semiclassical limit leads to eigenvalue statistics
the same as those of a suitable ensemble of random matrices.  For
instance, the probability distribution function for the separation
between neighboring eigenvalues is as derived from random matrix
theory and goes to zero at zero separation.  This contrasts with the
Poissonian distribution found in separable systems, showing that a
signature of quantum chaos is level repulsion.  In order to determine
whether eigenvalues of the regularized MHD problem obey the same
statistics as those of the Schr\"odinger equation in both the
separable 1-D case and the chaotic 3-D cases, we have assembled data
sets of ideal MHD eigenvalues for a Suydam-unstable cylindrical (1-D)
equilibrium using \emph{Mathematica} and a Mercier-unstable (3-D)
equilibrium using the CAS3D code.  In the 1-D case, we find that the
unregularized Suydam-approximation spectrum has an anomalous peak at
zero eigenvalue separation.  On the other hand, regularization by
restricting the domain of $\kvec_{\perp}$ recovers the expected
Poissonian distribution.  In the 3-D case we find strong evidence of
level repulsion within mode families, but mixing mode families
produces Poissonian statistics.  } \vspace*{0.20in}

%%%%%%%%    Keywords                                                 
%%%%%%%%


\vspace*{0.30in}

%\newpage

%-----------------------------------------------------------
]
\setlength{\baselineskip}{4.4mm}

%%%%%%%%    1.Introduction                                           
%%%%%%%%

\noindent
{\bf 1.Introduction}

% \section{Introduction\label{sec:Intro}}

In ideal MHD the spectrum of the growth rates, $\gamma$, of
instabilities is difficult to characterize mathematically because the
linearized force operator is not compact \cite{lifschitz89}.  This
gives rise to the possibility of a dense set of accumulation points
(descriptively called the ``accumulation continuum'' by Spies and
Tataronis \cite{spies-tataronis03} though more correctly termed
\cite{hameiri85} the \emph{essential spectrum}).

The continuous spectrum in quantum mechanics arises from the
unboundedness of configuration space, whereas the MHD essential
spectrum arises from the unboundedness of Fourier space---there is no
minimum wavelength in ideal MHD. This is an unphysical artifact of the
ideal MHD model because, in reality, low-frequency instabilities with
$|\kvec_{\perp}|$ much greater than the inverse of the ion Larmor radius,
$a_{\mathrm{i}}$, cannot exist (where $\kvec_{\perp}$ is the
projection of the local wavevector into the plane perpendicular to the
magnetic field $\Bvec$).

Perhaps the greatest virtue of ideal MHD in fusion plasma physics is
its mathematical tractability as a first-cut model for assessing the
stability of proposed fusion-relevant experiments with complicated
geometries.  For this purpose a substantial investment in effort has
been expended on developing numerical matrix eigenvalue programs, such
as the three-dimensional (3-D) TERPSICHORE \cite{anderson_etal90} and CAS3D
\cite{schwab93} codes.  These solve the MHD wave equations for
perturbations about static equilibria, so that the eigenvalue
$\omega^2 \equiv -\gamma^2$ is real due to the Hermiticity (self-adjointness
\cite{bernstein_etal58}) of the linearized force and kinetic energy
operators.  They use finite-element or finite-difference methods to
convert the infinite-dimensional Hilbert-space
eigenvalue problem to an
approximating finite-dimensional matrix problem.

In order properly to verify the convergence of these codes  in
3-D geometry it is essential to understand the nature of
the spectrum---if it is quantum-chaotic then convergence of individual
eigenvalues cannot be expected and a statistical description must be
used.

It is the thesis of this paper that the language of quantum chaos
\cite{haake01} theory indeed provides such a
statistical framework for characterizing MHD spectra in that it seeks to
classify spectra statistically by determining whether, and to what
degree, they belong to various universality classes. 

In the cylindrical case the eigenvalue problem is separable into three
one-dimensional (1-D) eigenvalue problems, with radial, poloidal, and
toroidal (axial) quantum numbers $l$, $m$, and $n$, respectively.  It
is thus to be expected \emph{a priori} that the spectrum will fall
within the standard quantum chaos theory universality class for
integrable, non-chaotic systems \cite{haake01}.  In particular, it is
to be expected that the probability distribution function for the
separation of neighboring eigenvalues is a Poisson distribution.
However, the nature of the MHD spectrum is quite different from that of
the typical quantum, microwave and acoustic systems normally dealt with
in quantum chaos theory and it is necessary to test this conjecture by
explicit calculation. In fact we find that the result depends on the
method of regularization.

We first present the eigenvalue equation for a reduced MHD model of a
large-aspect-ratio (effectively cylindrical) stellarator.  We study a
plasma in which the Suydam criterion \cite{suydam58} for the stability
of interchange modes is violated, so the number of unstable modes tends
to infinity as the small-wavelength cutoff tends to zero.  To compute
large-$m$ eigenvalues we transform to a Schr\"odinger-like form of the
radial eigenvalue equation \cite{cheremhykh-revenchuk92}, which has
essentially the same form in configuration ($r$) space as in Fourier
($k_r$) space, thus allowing easy regularization by restricting the
$k_r$ domain.  To simplify even further we approximate the effective
potential by a parabola, thus yielding the quantum harmonic oscillator
equation, solvable in parabolic cylinder functions
\cite{abramowitz-stegun65}.

Real, finite-aspect-ratio stellarators are fully 3-D and their
ideal-MHD spectra may be expected \emph{a priori} to fall within the
universality class appropriate to time-reversible quantum chaotic
systems, where the spectral statistics are found to be the same as for
a Gaussian orthogonal ensemble of random matrices \cite{haake01} in
regions where ray tracing reveals chaotic dynamics
\cite{dewar-cuthbert-ball01}.  At the end of this paper we give a brief
report of 3-D calculations peformed with the CAS3D code on a
Mercier-unstable, high-mirror-ratio, high-iota equilibrium representing
a Wendelstein 7-X (W7-X) stellarator variant \cite{nuehrenberg96}.

\vspace*{0.20in}
%%%%%%%%    2.One-dimensional model eigenvalue equation          %%%%%%%%

\noindent
{\bf 2.One-dimensional model eigenvalue equation}

In this paper we study an effectively circular-cylindrical MHD
equilibrium, using cylindrical coordinates such that the magnetic axis
coincides with the $z$-axis, made topologically toroidal by periodic
boundary conditions.  Thus $z$ and the toroidal angle $\zeta$ are
related through $\zeta \equiv z/R_0$, where $R_0$ is the major radius
of the toroidal plasma being modeled by this cylinder.  The poloidal
angle $\theta$ is the usual geometric cylindrical angle and the
distance $r$ from the magnetic axis labels the magnetic surfaces (the
equilibrium field being trivially integrable in this case).  The
plasma edge is at $r = a$.

In the cylinder there are two ignorable coordinates, $\theta$ and
$\zeta$, so the components of $\xivec$ are completely factorizable 
into
products of functions of the independent variables separately.  In
particular, we write the $r$-component as
\begin{equation}
		r\xi_r = \exp (im\theta )\exp (-in\zeta)\varphi(r) \;,
		\label{eq:sep}
\end{equation}
where the periodic boundary conditions quantize $m$ and $n$ to 
integers and we choose to work with the stream function 
$\varphi(r) \equiv r\xi_r(r)$.  

Since the primary motivation of this paper is stellarator physics, we
use the reduced MHD ordering for large-aspect stellarators
\cite{strauss80,wakatani98}, averaging over helical ripple to reduce
to an equivalent cylindrical problem
\cite{kulsrud63,tatsuno-wakatani-ichiguchi99}.  The universality class
should be insensitive to the precise choice of model as long as it
exhibits the behavior typical of MHD instabilities in a cylindrical
plasma, specifically the existence of interchange instabilities and
the occurrence of accumulation points at finite growth rates.

Defining $\lambda \equiv \omega^2$ we seek the spectrum of
$\lambda$-values satisfying the scalar equation
\begin{equation}
	 L \varphi = \lambda M\varphi
		\label{eq:eigvaleqn}
\end{equation}
under the boundary conditions $\varphi(0) = 0$ at the magnetic axis
and $\varphi(1) = 0$, appropriate to a perfectly conducting wall at
the plasma edge (using units such that $r=1$ there).

The operator $M = -\nabla_{\perp}^2$ and $L$ is given by
\begin{eqnarray}
	 L & = & -\frac{1}{r}\frac{d}{dr}(n - m\iotabar)^2 
r\frac{d}{dr} 
	 +\frac{m^2}{r^2}
	 	\left[
		(n - m\iotabar)^2 \right.
\nonumber \\  & & \left.\mbox{} 
          - \iotadot^2 G  +\frac{\iotaddot}{m}(n-m\iotabar)
	 	\right] \;,
 	 \label{eq:Ldef}
\end{eqnarray}
where $G$ is a Suydam stability parameter ($> 1/4$ for instability
\cite{suydam58}), proportional to the pressure gradient $p'(r)$ and the
average field line curvature \cite{wakatani98}.

In this paper we use the notation $\dot{f} \equiv
rf'(r)$ for an arbitrary function $f$, so $\iotadot \equiv rd\iota/dr$
is a measure of the magnetic shear and $\iotaddot$ measures the
variation of the shear with radius.

We observe some differences between \eq{eq:eigvaleqn} and the standard
quantum mechanical eigenvalue problem $H\psi = E\psi$.  One is of
course the physical interpretation of the eigenvalue---in quantum
mechanics the eigenvalue $E \equiv \hbar \omega$ is linear in the
frequency because the Schr\"odinger equation is first order in time,
whereas our eigenvalue $\lambda$ is quadratic in the frequency because
it derives from a classical equation of motion. 

Another difference is that \eq{eq:eigvaleqn} is a \emph{generalized}
eigenvalue equation because $M$ is not the identity operator.  This is
one reason why it is necessary to treat the MHD spectrum explicitly
rather than simply assume it is in the same universality class as
standard quantum mechanical systems.

Equation~(\ref{eq:eigvaleqn}) is very similar to the normal mode
equation analyzed in the early work on the interchange growth rate in
stellarators by Kulsrud \cite{kulsrud63}.  However, unlike this and
most other MHD studies we are concerned not with finding the highest
growth rate, but in characterizing the complete set of
unstable eigenvalues.

\begin{figure}[tbp]
\begin{center}
    \includegraphics[scale=0.7]{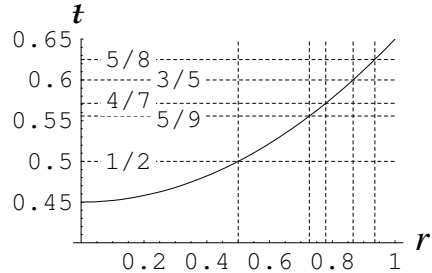}
\end{center}
    \caption{The rotational transform $\iotabar(r)\equiv 1/q(r)$
    with $\iotabar_0=0.45$,
    $\iotabar_2=0.2$.
    All distinct rational magnetic surfaces $\mu=n/m$ are shown for
    $m$ up to 10.}
    \label{fig:iprofiles}
\end{figure}

Suydam instabilities occur only for values of $m$ and $n$ such that $n
- m\iotabar$ vanishes.  For the 1-D numerical work in this paper we use a
parabolic transform profile $\iotabar = \iotabar_0 + \iotabar_2 r^2$ as
illustrated in \fig{fig:iprofiles}.

Given a rational fraction $\mu = n_{\mu}/m_{\mu}$ in the
interval $[\iotabar(0),\iotabar(a)]$ (where $n_{\mu} $and $m_{\mu}$
are mutually prime) there is a unique radius $\rmu$ such that
\begin{math}%\label{eq:rmudef}
    \iotabar(\rmu) = \mu \;.
\end{math}
Any pair of integers $ (m,n)_{\mu,\nu} \equiv (\nu m_{\mu}, \nu
n_{\mu})$, $\nu = 1, 2, 3, \ldots$ satisfies the resonance condition
\begin{equation}
    n_{\mu,\nu}  - m_{\mu,\nu} \iotabar(\rmu) = 0 \;.
    \label{eq:rescond}
\end{equation}

\begin{figure}[tbp] 
\begin{center}
    \includegraphics[scale=0.7]{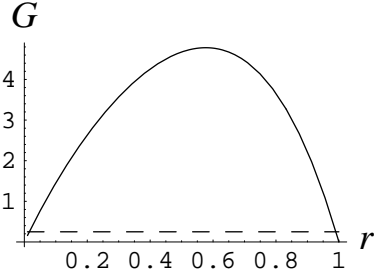} 
\end{center}
    \caption{The Suydam criterion parameter $G(r)$ (solid line), and
    the instability threshold $1/4$ (dashed line), showing nearly all
    the plasma is Suydam unstable.}
    \label{fig:Gprofile}
\end{figure}

We use a broad pressure profile that is sufficiently flat near the
magnetic axis that the Suydam instability parameter $G$ goes to zero at
the magnetic axis, and for which $p'$ vanishes at the plasma edge.
The resulting $G$-profile is shown in \fig{fig:Gprofile}.

Defining a scaled radial variable $x \equiv m(r-\rmu)/\rmu$, we can
find the large-$m$ spectrum of \eq{eq:eigvaleqn} by expanding all
quantities in inverse powers of $m$, and equating the LHS to zero order
by order.

In this paper we work only to lowest order in $1/m$, the \emph{Suydam 
approximation}.  As found by
Kulsrud \cite{kulsrud63}, we have the generalized eigenvalue equation
\begin{equation}
    \mathcal{L}^{(0)}\varphi^{(0)}
    \equiv
    \frac{\rmu^2}{m^2}\left(L^{(0)} - \lambda^{(0)}M^{(0)} \right)\varphi^{(0)}
    = 0 \;,
    \label{eq:eigvaleqn0}
\end{equation}
where, more explicitly,
\begin{equation}
    \frac{\mathcal{L}^{(0)}}{\iotadot^2}
	 = 
     -\frac{d}{dx}(x^2 + \Gamma^2) \frac{d}{dx}
     + x^2 + \Gamma^2 - G \;,
    \label{eq:calL0def}
\end{equation}
with $\Gamma^{2} \equiv -\lambda^{(0)}/\iotadot^2$ and $\iotadot$ and
$G$ evaluated at $\rmu$.  Under the boundary conditions $\varphi^{(0)}$
$\rightarrow$ $0$ as $r \rightarrow \pm\infty$, \eq{eq:eigvaleqn0} can
be solved to give a square-integrable eigenfunction, with growth rate
$\gamma = \iotadot \Gamma$, provided $\lambda^{(0)} < 0$ is one of the
eigenvalues $\lambda_{\mu,l}$.  The radial mode number $l =
0,1,2,\ldots$ denotes the number of nodes of the eigenfunction
$\varphi^{(0)} = \varphi_{\mu,l}(r)$.  Note that $\lambda_{\mu,l}$
depends only on $\mu = n/m$ and is otherwise independent of the
magnitude of $m$ and $n$.

Restricting attention to unstable modes, so that $\gamma \equiv
(-\lambda)^{1/2}$ is real, we transform \eq{eq:eigvaleqn0} to the
Schr\"odinger form \cite{cheremhykh-revenchuk92}
\begin{equation}
		 \frac{d^2\psi}{d\eta^2} + Q(\eta)\psi = 0 \;,
		\label{eq:Schrodinger}
\end{equation}
where
\begin{equation}
		Q \equiv  G - \quarter - \quarter\sech^2\,\eta
				       - \Gamma^2\cosh^2\eta  \;,
		\label{eq:Q0}
\end{equation}
with $\eta$ defined through $x \equiv
\gamma\sinh\eta/\iotadot(r_{\mu})$, and $\psi \equiv
(\cosh\eta)^{1/2}\varphi(x)$.

From, e.g., Eq.  (4.7) of \cite{cheremhykh-revenchuk92} we see that,
provided the Suydam criterion $G > 1/4$ is satisfied, there is an
infinity of $\gamma$ eigenvalues accumulating exponentially toward
the origin from above (so the $\lambda$-values accumulate from below)
in the limit $l \rightarrow \infty$.

Perhaps less widely appreciated (because $m$ and $n$ are normally
taken to be fixed) is the fact that there is also a point of
accumulation of the eigenvalues of \eq{eq:eigvaleqn} at each
$\lambda_{\mu,l}$ as $m \rightarrow \infty$ with $l$ fixed.
(Although $\lambda^{(0)}$ is infinitely degenerate, we can break this
degeneracy by proceeding further with the expansion in $1/m$, thus
showing that $\lambda_{\mu,l}$ is an accumulation point.)  Since the
rationals $\mu$ are dense on the real line, there is an ``accumulation
continuum'' \cite{spies-tataronis03} between $\gamma = 0$ and the
maximum growth rate, $\gamma = \gammamax$.

\vspace*{0.20in}
% %%%%%%%%    3.Regularization   %%%%%%%%
\noindent
{\bf 3.Regularization}

The accumulation points of the ideal MHD spectrum found above are
mathematically interesting but exist only as a singular limit of
equations containing more physics, including finite-Larmor-radius
(FLR) effects and dissipation, that regularize the spectrum.

In order to proceed further we need to be explicit about the nature of
this singular limit.  As we are primarily concerned with the
universality class question, we seek only a \emph{minimal}
modification of \eq{eq:eigvaleqn} that has some physical basis but
makes as little change to ideal MHD as possible.  To preserve the
Hermitian nature of ideal MHD we cannot use the drift correction used
for estimating FLR stabilization of interchange modes by Kulsrud
\cite{kulsrud63}.  However it is possible to effect a pseudo-FLR
regularization of ideal MHD by restricting $\kvec_{\perp}$ to a disk
of radius less than the inverse ion Larmor radius. 
In our nondimensionalized, large-aspect ratio model this implies
\begin{equation}
	(k_{\theta}^2 + k_r^2)^{1/2}\rho_{*} \leq 1 \;,
	\label{eq:cutoff}
\end{equation}
where $k_r$ and $k_{\theta}$ are the radial and poloidal components of
the wavevector, respectively, and $\rho_{*}$ is the ion Larmor radius (at
a typical energy) in units of the minor radius.

To apply \eq{eq:cutoff} precisely we need to relate $k_r$ and
$k_{\theta}$ to the eigenvalue problem discussed above.  From
\eq{eq:sep} we see that $k_{\theta} = m/r$.  We define $k_r$ as the
Fourier variable conjugate to $r$.  Fourier transformation of
\eq{eq:eigvaleqn} is only practical in the large-$m$ limit, when modes
are localized near the resonant surfaces $r = \rmu$, which is why we
have restricted the discussion to leading order in the $1/m$ expansion.

Using the stretched radial coordinate $x \equiv m(r-\rmu)/\rmu$ we
define $k_r \equiv m \kappa/\rmu$, where $\kappa$ is the
Fourier-space independent variable conjugate to $x$.
With the substitutions $d/dx \mapsto i\kappa$, $x \mapsto id/d\kappa$,
and using the fact that $\kappa d/d\kappa$ and $(d/d\kappa)\kappa
\equiv 1 + \kappa d/d\kappa$ commute, \eq{eq:eigvaleqn0} transforms to
\begin{equation}
		\left[
		 -\frac{d}{d\kappa}(1+\kappa^2) \frac{d}{d\kappa}  
		 + \Gamma^2(1+\kappa^2) - G
		\right]\varphi_{\kappa}  = 0\;.
		\label{eq:eigvaleqnk}
\end{equation}
The transformation $\kappa = \sinh\eta$ then leads back to
\eq{eq:Schrodinger}, with $\eta$ now to be interpreted as a distorted
Fourier-space independent variable, rather than as a real-space
coordinate!

Equation~(\ref{eq:cutoff}) implies that \eq{eq:eigvaleqnk} is to be 
solved on the domain $-\kappamax\leq \kappa \leq
\kappamax$ where
\begin{equation}
		\kappamax(\mu) \equiv 
		\left[\left(\frac{\rmu}{m\rho_{*}}\right)^2 - 1\right]^{1/2} \;.
		\label{eq:kappamax}
\end{equation}
This exists provided $|m| < \mmax$, where
\begin{equation}
		\mmax(\mu) \equiv \rmu/\rho_{*} \;.
\label{eq:mmax}
\end{equation}
Analogously to quantum mechanical box-quantization we use Dirichlet
boundary conditions at $\pm\kappamax$.

\vspace*{0.20in}
% %%%%%%%%    4.Spectral statistics in the 1-D case   %%%%%%%%
\noindent
{\bf 4.Spectral statistics in the 1-D case}

As only the qualitative nature of the spectrum is important, we
approximate the function $Q$ by $Q(0) + \half
Q''(0)\eta^{2}$, so \eq{eq:Schrodinger} can be solved in parabolic
cylinder functions \cite{abramowitz-stegun65}. We find the dispersion 
relation
\begin{equation}
		\nu + \half = 
		\frac{G - \Gamma^2 - \half}{(4\Gamma^2 - 1)^{1/2}}
		 \;,
		\label{eq:parabDR}
\end{equation}
where $\nu = l$ in the unregularized case, $\kappamax = \etamax =
\infty$.  In the even-$l$, regularized case $\nu$ may be found by
solving for a zero of $M(\nu/2, 1/2, (4\Gamma^2 -
1)^{1/2}\etamax^2/2)$, where $M$ is Kummer's function.  For $l=0$,
$\nu$ becomes exponentially small as $\etamax \rightarrow \infty$,
which allows an approximate regularization formula to be derived.

We study the spectrum between the maximum $l = 1$ growth rate,
$\gammamax(l=1) \equiv \max_{\mu}\gamma_{\mu,1}$, and the maximum
overall growth rate, $\gammamax = \max_{\mu}\gamma_{\mu,0}$.  Only
the $l=0$ modes exist in this range of $\gamma$, which corresponds to
the range in $\mu$ between $\mumin \approx 0.522$ and $\mumax \approx
0.628$. Throughout this range $\Gamma$ is $> 1/2$, so that $Q$ has a 
single minimum \cite{cheremhykh-revenchuk92} and the quadratic 
approximation of this section is appropriate. In this range there are 
only four low-order rationals $n/m$ with $m < 10$.

Taking $\rho_{*} = 0.001$, all pairs of integer values $m$, $n$ in the
fan-shaped region $1 \leq m \leq \mmax(n/m)$, $\mumin \leq n/m \leq
\mumax$ were evaluated, giving an initial dataset of over 32,000 points
$(m,n)$.  The corresponding set of unregularized eigenvalues was
calculated by solving \eq{eq:parabDR} with $\nu = 0$ and the
eigenvalues were sorted and numbered from the top to give the
integrated density of states ``staircase'' function $N(\gamma)$.

\begin{figure}[tbp]
\begin{center}
    \includegraphics[scale=0.7]{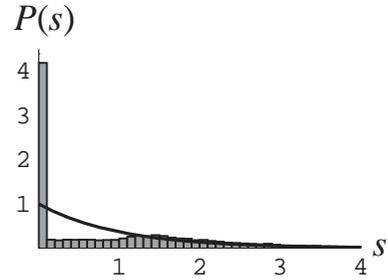}
\end{center}
    \caption{The histogram shows an estimate, based on a data set of
    about $32,000$ unregularized eigenvalues, of the probability
    distribution function for the eigenvalue separation $s$.  The
    plot is dominated by the spike at $s = 0$.}
    \label{fig:semiregPPlot}
\end{figure}

The curve $0.3523 - 9.5733\times 10^{-11}N^2 - 1.1625\times
10^{-20}N^4$ was found to give a good fit to the smoo\-th\-ed behavior of
this function.  Inverting this function gives the smoothed function
$\bar{N}(\gamma)$ which is used to ``unfold'' \cite{haake01}
spectra by defining a new ``energy eigenvalue'' $E \equiv
\bar{N}(\gamma)$, such that $N(E)$ increases linearly on average.

This means that the average separation of eigenvalues is now unity,
making comparison with spectra from other physical systems
meaningful and allowing universal behavior to become apparent if 
present.  However, \fig{fig:semiregPPlot} shows that the 
probability distribution of eigenvalue spacings $s$ is far from 
universal for the unregularized Suydam spectrum, exhibiting a 
delta-function-like spike at $s = 0$. This is presumably because, 
although we have truncated the spectrum in $m$, we have not removed 
the degeneracies arising for low-order rationals $\mu$ in the range 
$\mumin < \mu < \mumax$.

\begin{figure}[tbp]
\begin{center}
    \includegraphics[scale=0.7]{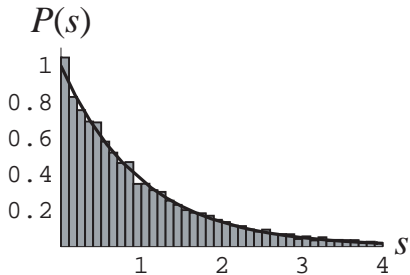}
\end{center}
    \caption{Eigenvalue spacing distribution for a data set of
    about $12,000$ regularized eigenvalues.  The
    exponential curve shows the Poisson distribution.}
    \label{fig:regPPlot}
\end{figure}

Figure~\ref{fig:regPPlot} on the other hand shows that when a similar
procedure is applied to the regularized spectrum (retaining only
regularized eigenvalues above $\gammamax(l=1)$, the universal Poisson
distribution expected from a separable system is obtained to
a good approximation, thus leading to the expectation that generic
quantum chaos theory is applicable once any physically reasonable
regularization is performed.

\begin{figure}[tbp]
\begin{center}
    \includegraphics[scale=0.4]{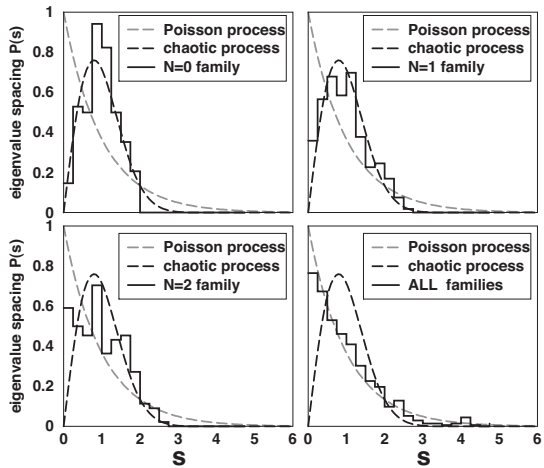}
\end{center}
    \caption{Eigenvalue spacing distributions from mode family datasets $N = 
    0$ (137 values), $N = 1$ (214 values) and $N = 2$ (178 values)
    from a W7-X-like equilibrium, and the distribution for the 
    combined spectrum, $N =0$, 1 and 2.}
    \label{fig:CASPPlot}
\end{figure}

Further support for this hypothesis is obtained from a CAS3D study of a
W7-X variant equilibrium with a nonmonotonic, low-shear transform
profile ($\iotabar_\mathrm{axis} = 1.1066$, $\iotabar_\mathrm{min} = 1.0491$,
$\iotabar_\mathrm{edge} = 1.0754$).  As seen from \fig{fig:CASPPlot}, when
the statistics are analyzed within the three mode families the
eigenvalue spacing distribution function is closer to the Wigner
conjecture form found for generic chaotic systems
\cite{haake01} than to the Poisson distribution for separable systems,
as might be expected from \cite{dewar-cuthbert-ball01}.  However, when the
spectra from the three uncoupled mode familes are combined, there are
enough accidental degeneracies that the spacing distribution becomes
close to Poissonian.

\vspace*{0.20in}

\vspace*{0.20in}
%%%%%%%%    Acknowledgments                                 %%%%%%%%

\noindent
{\bf Acknowledgments}:
Part of this work was performed at the University of Tokyo,
Graduate School of Frontier Sciences (RLD,TT).
We thank 
Professor Zensho Yoshida for his hospitality, support and useful 
discussions. RLD was partially
supported by the Australian Research Council.

\vspace*{0.20in}

%%%%%%%%    References                                               
%%%%%%%%

\noindent
{\bf References}

\noindent
\bibliography{Ballooning}% Produces the bibliography via BibTeX.

\end{document}